\documentclass{esannV2}
\usepackage[dvips]{graphicx}
\usepackage[latin1]{inputenc}
\usepackage{amssymb,amsmath,array}
\usepackage{floatrow}

\begin{document}
\title{An objective function for self-limiting neural plasticity rules.}

\author{Rodrigo Echeveste and Claudius Gros
%
%
\vspace{.3cm}\\
%
Institute for Theoretical Physics - Goethe University Frankfurt\\
Frankfurt am Main - Germany
}

\maketitle

\begin{abstract}
Self-organization provides a framework for the study of systems 
in which complex patterns emerge from simple rules, without the 
guidance of external agents or fine tuning of parameters. 
Within this framework, one can formulate a guiding principle for 
plasticity in the context of unsupervised learning, in 
terms of an objective function. In this work we derive Hebbian, 
self-limiting synaptic plasticity rules from such an objective 
function and then apply the rules to the non-linear bars problem.
\end{abstract}

\section{Introduction}

Hebbian learning rules \cite{hebb2002organization} are
at the basis of unsupervised learning in neural networks,
involving the adaption of the inter-neural synaptic weights 
\cite{bienenstock1982theory,oja1997nonlinear}. These rules 
usually make use of either an additional renormalization 
step or a decay term in order to avoid runaway synaptic 
growth \cite{goodhill1994role,elliott2003analysis}.\\

From the perspective of self-organization 
\cite{kohonen1988self,gros2010complex,gros2014generating,nicolis1977self}, 
it is interesting 
to study how Hebbian, self-limiting synaptic plasticity rules 
can emerge from a set of governing principles, in terms of 
objective functions. Information theoretical measures such 
as the entropy of the output firing rate distribution have 
been used in the past to generate rules for either intrinsic 
or synaptic plasticity 
\cite{triesch2007synergies,stemmler1999voltage, markovic2012intrinsic}. 
The objective function with which we work here can be motivated 
from the Fisher information, which measures the sensitivity 
of a certain probability distribution to a parameter, in this 
case defined with respect to the Synaptic Flux operator 
\cite{echeveste2014generating}, 
which measures the overall increase of synaptic weights. 
Minimizing the Fisher information corresponds, in this 
context, to looking for a steady state solution where the 
output probability distribution is insensitive to local 
changes in the synaptic weights. This method, then 
constitutes an implementation of the stationarity 
principle, stating that once the features of a stationary 
input distribution have been acquired, learning should stop, 
avoiding runaway growth of the synaptic weights.\\

It is important to note that, while in other contexts 
the Fisher information is maximized to estimate 
a certain parameter via the Cram\'er-Rao bound, in this 
case the Fisher information is defined with respect to 
the model's parameters, which do not need to be estimated, 
but rather adjusted to achieve a certain goal. This 
procedure has been successfully employed in the past in 
other fields to derive, for instance, the Schr\"odinger 
Equation in Quantum Mechanics 
\cite{reginatto1998derivation}.\\

\section{Methods}

We consider rate-encoding point neurons, where the output 
activity $y$ of each neuron is a sigmoidal 
function of its weighed inputs, as defined by:
\begin{equation}
y = g(x), \qquad
x=\sum_{j=1}^{N_w} w_j (y_j-\bar y_j).
\label{eq_neuron_model}
\end{equation}
Here the $y_js$ are the $N_w$ inputs to the neuron (which 
will be either the outputs of other neurons or external 
stimuli), the $w_j$ are the synaptic weights, and $x$ 
the integrated input, which one may consider as the 
neuron's membrane potential. $\bar y_j$ represents the average 
of input $y_j$, so that only deviations from the average 
convey information. $g$ represents here a sigmoidal 
transfer function, such that $g(x)\longrightarrow1/0$ 
when $x\longrightarrow \pm \infty$. The output firing rate $y$ 
of the neuron is hence a sigmoidal function of the membrane 
potential $x$.\\

By minimization through stochastic gradient descent of:
\begin{equation}
F_{ob} \ =\ E\big[f_{ob}(x)\big] \ =\ E\left[\big(N+ A(x)\big)^2\right],
\qquad A(x) = \frac{xy''}{y'}~,
\label{eq_F_ob}
\end{equation}
a Hebbian self-limiting learning rule for the synaptic weights 
can be obtained \cite{echeveste2014generating}.
Here $E[.]$ denotes the expected value, as averaged over 
the probability distribution of the inputs, and $y'$ and 
$y''$ are respectively the first and second derivatives of 
$y(x)$. $N$ is a parameter of the model (originally derived as 
$N_w$ and then generalized \cite{echeveste2014generating}), 
which sets the values for the system's fixed-points, as 
shown in Section \ref{sec_minima}.

In the case of an exponential, or Fermi transfer function,
we obtain
\begin{equation}
g_{exp}(x) \ =\ \frac{1}{1+exp(b-x)},
\qquad\quad
f_{ob} \ =\ \Big(N+x\big(1-2y(x)\big)\Big)^2
\label{eq_sigma_expo}
\end{equation}
for the kernel $f_{ob}$ of the objective function 
$F_{ob}$. The intrinsic parameter $b$ represents a bias and 
sets the average activity level of the neuron. This parameter 
can either be kept constant, or adapted with little 
interference by other standard procedures such as 
maximizing the output entropy 
\cite{triesch2007synergies,echeveste2014generating}.\\

\begin{figure}[t]
\centering
\includegraphics[width=0.9\textwidth]{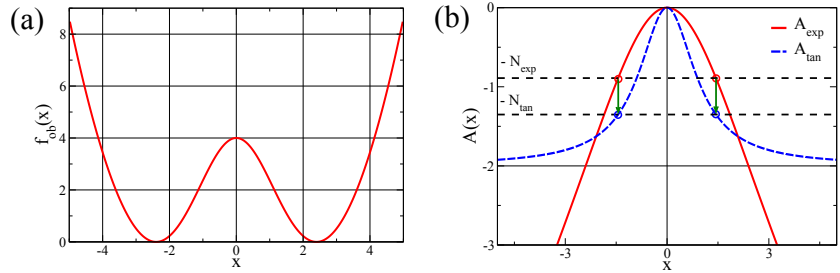}
\caption{\textbf{(a)} $f_{ob}(x)$, as defined
by Eq.~(\ref{eq_sigma_expo}), for $b=0$ and $N=2$. 
The synaptic weights are adapted through (\ref{eq_w_dot})
such that the membrane potential $x$ tends to cluster
around the two minima.
\textbf{(b)} $A(x)$, as defined by Eqs.~(\ref{eq_F_ob}) and
(\ref{eq_As}), for both the exponential and the tangential 
sigmoidal transfer functions and $b=0$. Adapting the
respective values of $N$ identical roots can be obtained,
as illustrated graphically.
}\label{Fig_f_ob}
\end{figure}

In  Fig.~\ref{Fig_f_ob}\textbf{(a)} the functional dependence of 
$f_{ob}$ is shown. It diverges for $x\longrightarrow \pm \infty$
and minimizing $f_{ob}$ will hence keep $x$, and therefore 
the synaptic weights, bound to finite values.
Minimizing (\ref{eq_sigma_expo}) through stochastic 
gradient descent with respect to $w_j$, one obtains 
\cite{echeveste2014generating}:
\begin{eqnarray}
\label{eq_w_dot}
&\dot w_j \ =\ \epsilon_w G(x)H(x)(y_j-\bar{y}_j) &\\
&G(x) = N+x(1-2y), \qquad \quad H(x) = (2y-1)+2x(1-y)y 
&\label{eq_G_H}
\end{eqnarray}
where the product $H(x)(y_j-\bar{y}_j)$ represents the Hebbian 
part of the update rule, with $H$ being an increasing function 
of $x$ or $y$, and where $G$ reverses the sign when the activity
is too large to avoid runaway synaptic growth. 

\subsection{Minima of the objective function}\label{sec_minima}

While (\ref{eq_F_ob}) depends quantitatively on the 
specific choice of the transfer function $g$, we will 
show here how the resulting expression for different transfer 
functions are in the end similar. We compare here as an example 
two choices for $g$, the exponential sigmoidal (or Fermi function) 
defined in (\ref{eq_sigma_expo}), and a arc-tangential 
transfer function defined as:
\begin{equation}
g_{tan}(x) \ =\ \frac{1}{\pi}arctan(x-b) +1/2.
\label{eq_sigma_tan}
\end{equation}
These two choices of $g$, in turn, define two versions of 
$A(x)$,
\begin{equation}
A_{exp}(x) \,=\,x\big(1-2y(x)\big)
\quad\qquad A_{tan}(x) \,=\,-\frac{2x(x-b)}{1+(x-b)^2}.
\label{eq_As}
\end{equation}
The objective functions are strictly positive $f_{ob}\geq0$, 
compare (\ref{eq_F_ob}), and their roots 
\begin{equation}
A_{exp/tan}(x) \ =\ -N
\label{roots}
\end{equation}
correspond hence to global minima, which are illustrated 
in Fig.~\ref{Fig_f_ob}\textbf{(b)}, where $A_{exp}(x)$ and $A_{tan}(x)$ 
are plotted for $b=0$. The minima of $f_{ob}$ can then be easily 
found by the intersection of the plot of $A(x)$ with the 
horizontals at $-N$. For $A_{exp}(x)$ one finds global minima 
for all values of $N$, whereas $N$ needs to be within
$[0,2]$ for the case of $A_{tan}(x)$. $N$ is however just a parameter 
of the model and the roots of the function which correspond to the 
neuron's membrane potential are in the same range, with each root 
representing a low- and high activity states.\\

While both rules display a similar behavior, they are not identical. 
$f_{ob}$ diverges for $x \longrightarrow \pm \infty$ keeping the 
weights bound, regardless of the dispersion in the input 
distribution. The maxima for $x \longrightarrow \pm \infty$ in the 
tangential function are of finite height, and this height decreases 
with $N$, making it unstable to noisy input distributions for larger 
values of $N$.\\

\subsection{Applications: PCA and the non-linear bars problem}

In \cite{echeveste2014generating}, the authors showed how a 
neuron operating under these rules is able to find the first 
principal component (PC) of an ellipsoidal input distribution. 
Here we present the neuron with Gaussian activity distributions 
$p(y_j)$ (the distributions are truncated so that $y_j\in[0,1]$). 
A single component, in this case $y_1$, has standard 
deviation $\sigma$ and all other $N_w-1$ directions have a 
smaller standard deviation of $\sigma/2$ (the rules are, however, 
completely rotation invariant). As an example, we have 
taken $N_w=100$, and show how with both transfer functions, the 
neuron is able to find the PC.\\
\begin{figure}[t]
\floatbox[{\capbeside\thisfloatsetup{capbesideposition={left,top},
capbesidewidth=0.4\textwidth}}]{figure}[\FBwidth]
{\caption{Evolution of the synaptic weights for both transfer 
functions (\ref{eq_sigma_expo}) and (\ref{eq_sigma_tan}). 
The continuous line represents $w_1$, corresponding 
to the principal component. A representative subset of 
the $N_w-1=99$ other weights is presented as dotted lines.
Top: exponential transfer function. Bottom: tangential 
transfer function.}\label{fig_weights_exp_tan}}
{\includegraphics[width=0.55\textwidth]{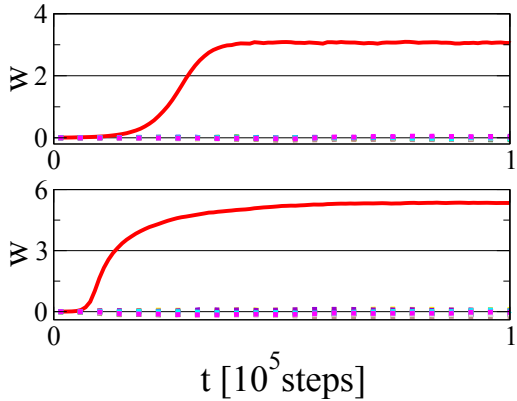}}
\end{figure}

In Fig.~\ref{fig_weights_exp_tan}, the evolution of the synaptic 
weights is presented as a function of time. In this case $b$ has 
been kept constant at $b = 0$. Learning stops when $<\dot w> = 0$, 
but since the learning rule is a non-linear function of $x$, the 
exact final value of $w$ will vary for different transfer 
functions. In the case of a bimodal input distribution, as the one used 
in the linear discrimination task, both clouds of points can 
be sent close to the minima and the final values of $w$ are then
very similar, regardless of the choice of transfer function (not 
shown here).\\

Finally, we apply the rules to the non-linear bars problem.  
Here we follow the procedure of \cite{foldiak1990forming}, 
where, in a grid of $L\times L$ inputs, each pixel can take 
two values, one for low intensity and one of high intensity. 
Each bar consists of a complete row or a complete column of 
high intensity pixels, and each possible bar is drawn 
independently with a probability $p=1/L$. At the intersection 
of a horizontal and vertical bar, the intensity is the same as 
if only one bar were present, which makes the problem non-linear.
The neuron is then presented, at each training step, a new 
input drawn under the prescribed rules and after each step 
the evolution of the synaptic weights is updated. The bias 
$b$ in the model can either be adjusted as in 
\cite{foldiak1990forming} by, $\dot{b} \propto (y-p)$, or 
by maximal entropy intrinsic adaption, as described in 
\cite{echeveste2014generating}, without mayor differences.\\

\begin{figure}[t]
\centering
\includegraphics[width=0.9\textwidth]{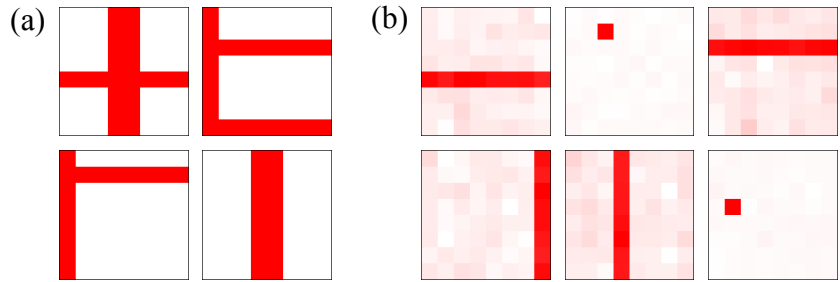}
\caption{\textbf{(a)} Some random training examples for 
the non-linear bars problem. \textbf{(b)} Graphical 
representation of the typical weight vectors learnt 
by the neuron in distinct training runs.
} \label{Fig_bars}
\end{figure}

Since the selectivity to a given pattern is given by the value 
of the scalar product $\bar y_{inputs} \cdot \bar w$, one can 
either compute the output activity $y$ to see to which pattern 
the neuron is selective in the end, or just do an intensity plot 
of the weights, since the maximal selectivity corresponds to 
$\bar w \propto \bar y_{inputs}$. In Fig.~\ref{Fig_bars} a 
typical set of inputs is 
presented, together with a typical set of learnt neural weights 
for different realizations in a single neuron training. We 
see how a neuron is able to become selective to individual 
bars or to single points (the independent components in 
this problem). To check that the neuron can learn single bars, 
even when such a bar is never presented to the neuron in 
isolation as a stimulus, we also trained the neuron with a 
random pair of bars, one horizontal and one vertical, 
obtaining similar results. The neuron can learn to fire 
in response to a single bar, even when that bar was never 
presented in isolation.\\

\section{Discussion and Concluding Remarks}

The implementation of the stationarity principle in terms of 
the Fisher information, presented in \cite{echeveste2014generating}
and here discussed, results in a set of Hebbian self-limiting 
rules for synaptic plasticity. The sensitivity of the rule to 
higher moments of the input probability distribution, makes it 
suitable for applications in independent component analysis.
Furthermore, the learning rule derived is robust with respect to 
the choice of transfer function $g(x)$, a requirement for biological
plausibility.\\

In upcoming work, we study the dependence of the steady state 
solutions of the neuron and their stability with respect to the 
moments of the input distribution. The numerical finding 
of independent component analysis in the bars problem is then justified. 
We will also study how a network of neurons can be trained using the 
same rules for all weights, feed-forward and 
lateral, and how clusters of input selectivity to different 
bars emerge in a self organized way.\\


\begin{footnotesize}

\bibliographystyle{unsrt}
\bibliography{Flux_Esann.bib}

\end{footnotesize}


\end{document}